\documentclass{article}
\usepackage{graphicx}
\usepackage{amssymb}
\usepackage{amsmath}
\usepackage{cite}

\begin{document}

\title{On the Entanglement Entropy of Maxwell Theory: \\
A Condensed Matter Perspective}
\author{Michael Pretko \\
\emph{Department of Physics and Center for Theory of Quantum Matter} \\
\emph{University of Colorado, Boulder, CO 80309}}
\date{\today}
\maketitle

\large

\begin{abstract}
Despite the seeming simplicity of the theory, calculating (and even defining) entanglement entropy for the Maxwell theory of a $U(1)$ gauge field in (3+1) dimensions has been the subject of controversy.  It is generally accepted that the ground state entanglement entropy for a region of linear size $L$ behaves as an area law with a subleading logarithm, $S = \alpha L^2 -\gamma \log L$.  While the logarithmic coefficient $\gamma$ is believed to be universal, there has been disagreement about its precise value.  After carefully accounting for subtle boundary corrections, multiple analyses in the high energy literature have converged on an answer related to the conformal trace anomaly, which is only sensitive to the local curvature of the partition.  In contrast, a condensed matter treatment of the problem yielded a topological contribution which is not captured by the conformal field theory calculation.  In this perspective piece, we review aspects of the various calculations and discuss the resolution of the discrepancy, emphasizing the important role played by charged states (the ``extended Hilbert space") in defining entanglement for a gauge theory.  While the trace anomaly result is sufficient for a strictly pure gauge field, coupling the gauge field to dynamical charges of mass $m$ gives a topological contribution to $\gamma$ which survives even in the $m\rightarrow\infty$ limit.  For many situations, the topological contribution from dynamical charges is physically meaningful and should be taken into account.  We also comment on other common issues of entanglement in gauge theories, such as entanglement distillation, algebraic definitions of entanglement, and gauge-fixing procedures.
\end{abstract}

\tableofcontents

\newpage

\section{Introduction}

In recent years, the concept of entanglement has become an increasingly important tool for characterizing quantum systems with many degrees of freedom, for both the condensed matter and high energy communities.  For example, many quantum systems, such as topological phases of matter, can be characterized by the inability of their ground state wavefunction to be smoothly disentangled to a direct product state \cite{topo}.  However, the entanglement pattern of a many-body wavefunction contains far too much information to make sense of directly.  In order to extract anything useful, we must boil the entanglement down to some simple metric.  There are numerous entanglement metrics available, such as Renyi entropies, mutual information\cite{mutual}, entanglement negativity\cite{neg}, etc.  But it seems fair to say that, at the present time, the traditional von Neumann entanglement entropy has proven to be the most useful characterization of entanglement.  This quantity is fairly straightforward to define, at first glance.  We begin with a pure state $|\Psi\rangle$ of the entire system, which we partition into two spatial\footnote{There have also been some investigations into the entanglement associated with partitions in momentum space\cite{momentum}, but to date, it is real space partitions which have proven the most informative.} subregions, $A$ and $B$.  We define the entanglement entropy of the partition to be the von Neumann entropy of the reduced density matrix for one subregion:
\begin{equation}
S = -Tr_A [\rho_A\log\rho_A]
\end{equation}
where $\rho_A = Tr_B[|\Psi\rangle\langle\Psi|]$.  It is easy to show that $\rho_A$ and $\rho_B$ have the same entropy, so it does not matter which subregion we examine.

This definition seems simple enough, but there is an important subtlety encountered in the case of gauge theories.  In order to speak sensibly about a spatial partition of our system, our theory must be described in terms of local degrees of freedom.  More formally, our system must have a tensor product Hilbert space.  The states of a pure gauge field, however, do not admit such a tensor product decomposition.  Luckily, there is a simple physically-motivated solution to this conundrum.  While a pure gauge field does not have a tensor product Hilbert space, a gauge field coupled to charged matter \emph{does} have such a structure, as we will review.  Thus, as long as we account appropriately for the charge sector of the Hilbert space, the definition of entanglement entropy will extend naturally to the case of gauge fields.

The entanglement entropy contains far less information than the full ground state wavefunction.  Nevertheless, entanglement entropy remains informative enough to yield several important insights into the study of quantum field theories.  First of all, it teaches us that the properties of ground states are far from generic within the Hilbert space.  With a few notable exceptions\cite{1dcft1,1dcft2,fermi1,fermi2,fermi3,ramis}, ground states of local quantum field theories obey an ``area law" for entanglement entropy\cite{area}.  For a spatial partition of linear size $L$ in $d$ spatial dimensions, the leading term of the entanglement entropy for a ground state is proportional to the surface area of the partition, $S\sim L^{d-1}$.  In contrast, a generic state of the Hilbert space would obey a ``volume law," $S\sim L^d$.  (Similar area law behavior is also seen in the context of many-body localization \cite{rahul}, and has intriguing connections with image processing via neural networks \cite{yahui}.)  While the existence of a ground state area law is fairly universal, its coefficient is not, depending sensitively on the small-scale regularization of the theory.  The more interesting aspects of entanglement entropy are actually encoded in the subleading terms to the area law.  For example, topologically ordered phases of matter in $(2+1)$ dimensions, which are well-described by topological field theories, are characterized by a subleading constant in the ground state entanglement entropy: $S = \alpha L-\gamma$.  The first term is the area law, which has a non-universal coefficient, determined by the lattice regularization.  The subleading constant $\gamma$, however, is topological in origin and is independent of the lattice, serving as a universal characterization of the phase.\footnote{It cannot, however, always uniquely identify a phase.  For example, the toric code and double semion models both have $\gamma = \log 2$, so one must resort to more complicated methods to distinguish these states.}  Furthermore, special techniques have been designed to isolate this ``topological entanglement entropy," discarding the area law and any other terms which are only sensitive to local physics\cite{kitaev,levin,grover}.

Given that subleading terms of entanglement entropy are powerful tools in characterizing topological field theories, including Chern-Simons gauge theories, it seems natural to also study the entanglement entropy of an even more familiar gauge theory, the simple Maxwell field theory of a $U(1)$ gauge field in $(3+1)$ dimensions.  The problem of entanglement in Maxwell theory has been intensely and independently studied by numerous groups over the past decade \cite{solo,buiv1,buiv2,chm,dowker, casini,numercas,revisit,janet1,instanton,nozaki,disk,donabelian,don1,kuo,ghosh,aoki,don2,radi,me,aspect,cas3,zuo,trivedi,agarwal,lin}.\footnote{Apologies are due in advance for any missed citations within the high energy literature, which is not the primary field of the present author.  Readers are encouraged to share any other relevant literature, to be added in a future version.}  It is by now generally agreed upon that the ground state entanglement entropy for a region of linear size $L$ takes the form of an area law with a universal logarithmic correction:
\begin{equation}
S = \alpha L^2 - \gamma \log L
\end{equation}
The area law coefficient $\alpha$ is non-universal, as always.  On the other hand, the dimensionless logarithmic coefficient $\gamma$ is insensitive to details of lattice regularization.  But while there is wide agreement that the $\gamma$ coefficient is universal, there has been significant disagreement regarding its precise value.  We briefly summarize some of the results that have been obtained.

\subsection*{The High Energy Literature}

Some of the earliest work\cite{solo,chm} on this problem relied on the fact that pure $(3+1)$-dimensional Maxwell theory is conformally invariant.  A simple argument (which we will review) indicates that a subleading logarithm should be present in the entanglement entropy for any $(3+1)$-dimensional conformal field theory, with a coefficient given in terms of the trace anomaly.  For Maxwell theory with a spherical partition, this yields a coefficient of $\gamma = 31/45$.  A subsequent calculation by Dowker\cite{dowker} based on thermodynamic arguments seemed to call this result into question.  By calculating the entropy of a particular thermal distribution of photons, Dowker obtained the smaller coefficient of $16/45$.  Recently, however, multiple groups in the high energy community have identified a missing ``boundary term" of $1/3$ in the thermodynamic calculation which precisely compensates for the difference \cite{don1,kuo,don2,zuo,trivedi}.  The sum of the bulk and boundary contributions once again yields the trace anomaly result for the total entanglement entropy:
\begin{equation}
\gamma_{H.E.} = \frac{16}{45} + \frac{1}{3} = \frac{31}{45}
\label{he}
\end{equation}
We refer to this as ``the" high energy result for the total logarithmic coefficient for a spherical partition.  There are, however, some conflicting perspectives within the high energy community, which we will address in Section 4.

There are two important features of this high energy result which are worth noting.  First of all, this result is determined purely in terms of the local curvature of the partition and is insensitive to its global topology.  This result is therefore ``topologically trivial," in a sense we will make precise.  Second, a distinction is often made between the two terms of the sum in Equation \ref{he}.  The $16/45$ term, coming from the bulk, is ``extractable" via entanglement distillation, $i.e.$ it can be removed from the system by physical operations on the gauge mode \cite{trivedi}.  In contrast, the boundary $1/3$ term is ``non-extractable" via such operations and is often regarded as unphysical.  We will return to discuss this issue further in Section 4, where we emphasize that this ``non-extractable" piece of the entanglement entropy is still an important physical quantity, corresponding to entanglement which can only be removed via operations on the charge sector of the theory.

\subsection*{The Condensed Matter Result}

Meanwhile, in a different corner of the physics universe, the problem of entanglement in Maxwell theory was independently studied in the condensed matter literature, by the present author and T. Senthil\cite{me}.  The problem is of direct interest to condensed matter physicists since Maxwell theory provides the appropriate low-energy theory for the simplest $U(1)$ spin liquid, which is a promising candidate for a real spin liquid in certain pyrochlore materials\cite{bosfrc3d,hfb04,3ddmr,lesikts05,kdybk,levinu1f,balentsqspice}.  Such systems feature an emergent $U(1)$ gauge field coupled to massive emergent charges, and at sufficiently low energy, Maxwell theory provides an adequate description for most of the physics.  In our previous work, we used a thermodynamic method similar to that used by Dowker to calculate the entanglement entropy.  We also found that the logarithmic coefficient breaks up into two pieces, one coming from the bulk and one from the boundary.  The bulk term arises from the same thermal distribution of photons considered by Dowker, giving a contribution of $16/45$ for a spherical partition.  (We misquoted this result in the first version of Reference \cite{me}.  See the note added in proof therein.)

The essential difference between the condensed matter result and the high energy literature occurs at the level of the boundary term.  In contrast to the high energy result, we found a boundary contribution given by a topological invariant.  Specifically, the boundary term is given by the number of connected components (zeroth Betti number) of the partitioning surface.  Combined with the bulk term, the total condensed matter answer for the entanglement entropy for a spherical partition is:
\begin{equation}
\gamma_{C.M.} = \frac{16}{45} + 1
\end{equation}
We intentionally keep the two terms separate to emphasize their different behavior.  The first term is sensitive only to the local curvature of the partition.  The second term, on the other hand, is a topological invariant, independent of the precise shape of the partition.  Furthermore, as we will review, a special procedure can be applied to isolate the topological contribution, eliminating the local term.  The $1$ represents the most robust characterization of Maxwell theory and is therefore the most interesting piece of the entanglement entropy, while the $16/45$ is comparatively boring.  This topological contribution can be seen explicitly in certain trial wavefunctions for $U(1)$ spin liquids\cite{me,wen2}\footnote{The result of Reference \cite{wen2} is in (2+1) dimensions, where the topological piece becomes $1/2$ instead of $1$.  More generally, the topological contribution to $\gamma$ in $(d+1)$ dimensions takes the form $\gamma_{top} = (d-1)/2$.} and has a simple physical interpretation in terms of charge neutrality of the ground state wavefunction.  The fact that calculations in the high energy literature did not yield such a contribution has therefore been somewhat puzzling.

\subsection*{The Resolution}

In the present work, we analyze and explain the discrepancy between the high energy and condensed matter results.  As we discuss, the boundary term represents a contribution from the charge sector of the theory, corresponding to a thermal distribution of charge living on the boundary.  This boundary term is highly sensitive to the precise way in which we treat charges.  Indeed, sensitivity of gauge theory entanglement to the presence of charges has been noted previously\cite{matter}.  In strictly pure Maxwell theory, the boundary term involves only non-dynamical ``test charges," which have no ability to react and respond to each other.  In contrast, when we include dynamical charges in the Hilbert space, the thermal boundary charges acquire the ability to \emph{screen} each other.  The resulting screened Coulomb gas has only short-range correlations, in contrast with the bare long-range correlations of the unscreened test charges.  This sudden change in the behavior of correlation functions upon the introduction of charge dynamics manifests itself in a very different contribution to the entanglement entropy.

The contribution of dynamical charges to the entanglement entropy is different from the high energy result in one other important aspect.  Note that the argument of a logarithm should technically be dimensionless, so the partition size $L$ must be compared against some other length scale in the argument of the subleading logarithm.  In the conformal field theory calculation, the only scale in the problem is the short-distance cutoff, $\epsilon$, so the logarithm must have the form $\log(L/\epsilon)$.  In contrast, dynamical charges lead to a logarithmic contribution to the entanglement entropy of the form $\log(L/m\epsilon^2)$, where $m$ is the mass scale of the charges.  By accounting for dynamical charges, a dimensionful parameter has been introduced into the problem, modifying the results of the conformal field theory calculation.  We see that, in this sense, the conformal limit of Maxwell theory is a somewhat singular one.  For any finite particle mass, there are logarithmic terms in the entanglement entropy which are not captured by the pure conformal field theory.

For a given physical situation, one must carefully consider what is truly meant by ``Maxwell theory."  If we really mean a strictly isolated gauge field, with no charges at any energy scale, then the trace anomaly result will hold.  However, if Maxwell theory is obtained as a low-energy theory by integrating out massive dynamical charges, then the topological result will hold.  For essentially all condensed matter purposes, such as the $U(1)$ spin liquid, and for many high energy purposes, such as the Standard Model, this latter case is the physically relevant one.  However, the former result may also have application within certain contexts.

\section{The Hilbert Space}

\begin{figure}[b!]
 \begin{minipage}[b]{0.45\linewidth}
 \centering
 \includegraphics[scale=0.3]{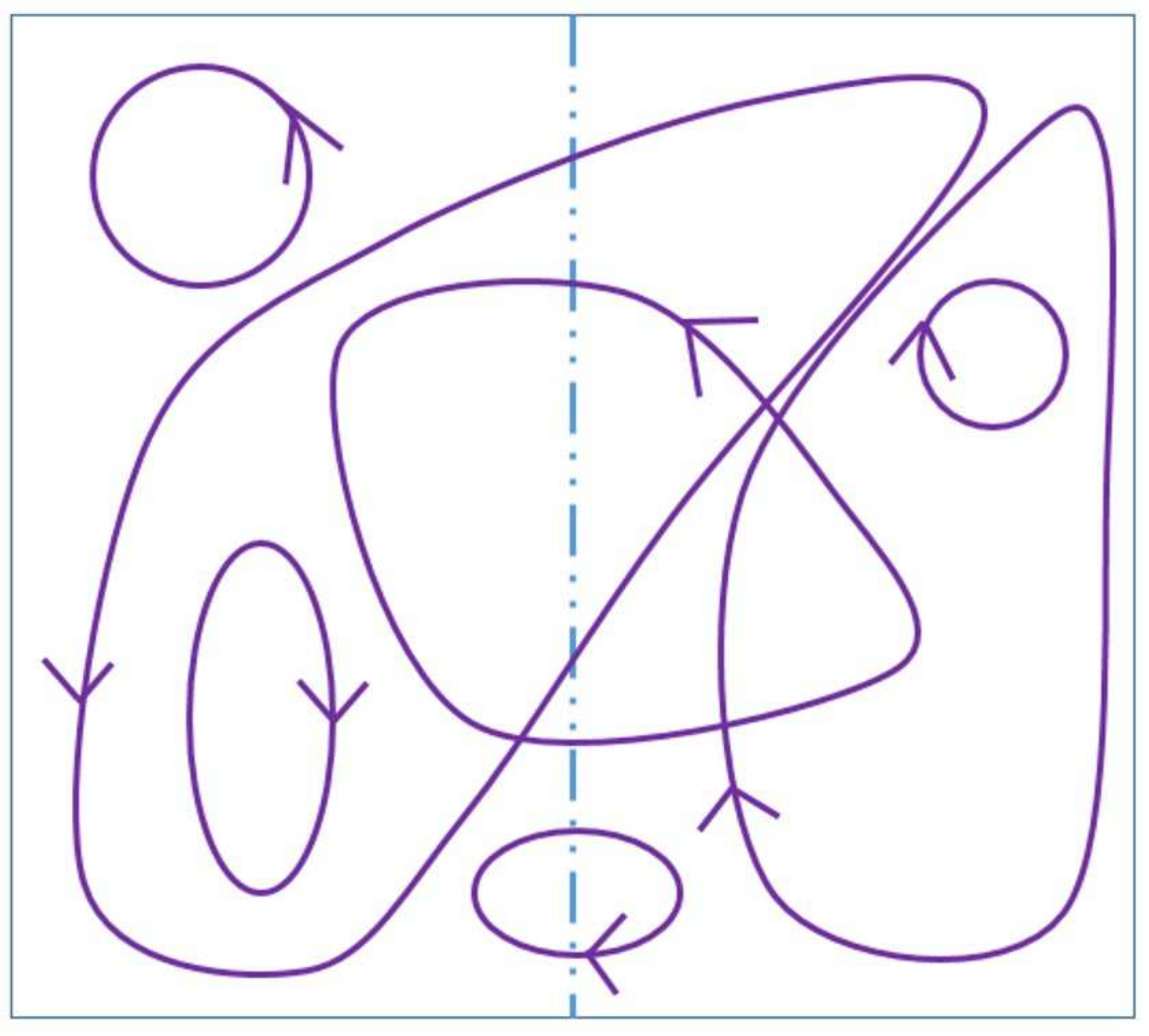}
 \caption{The pure photon states are superpositions of closed loop configurations of the electric field, such as seen above.}
 \label{fig:closed}
 \end{minipage}
 \hspace{1cm}
 \begin{minipage}[b]{0.45\linewidth}
 \centering
 \includegraphics[scale=0.3]{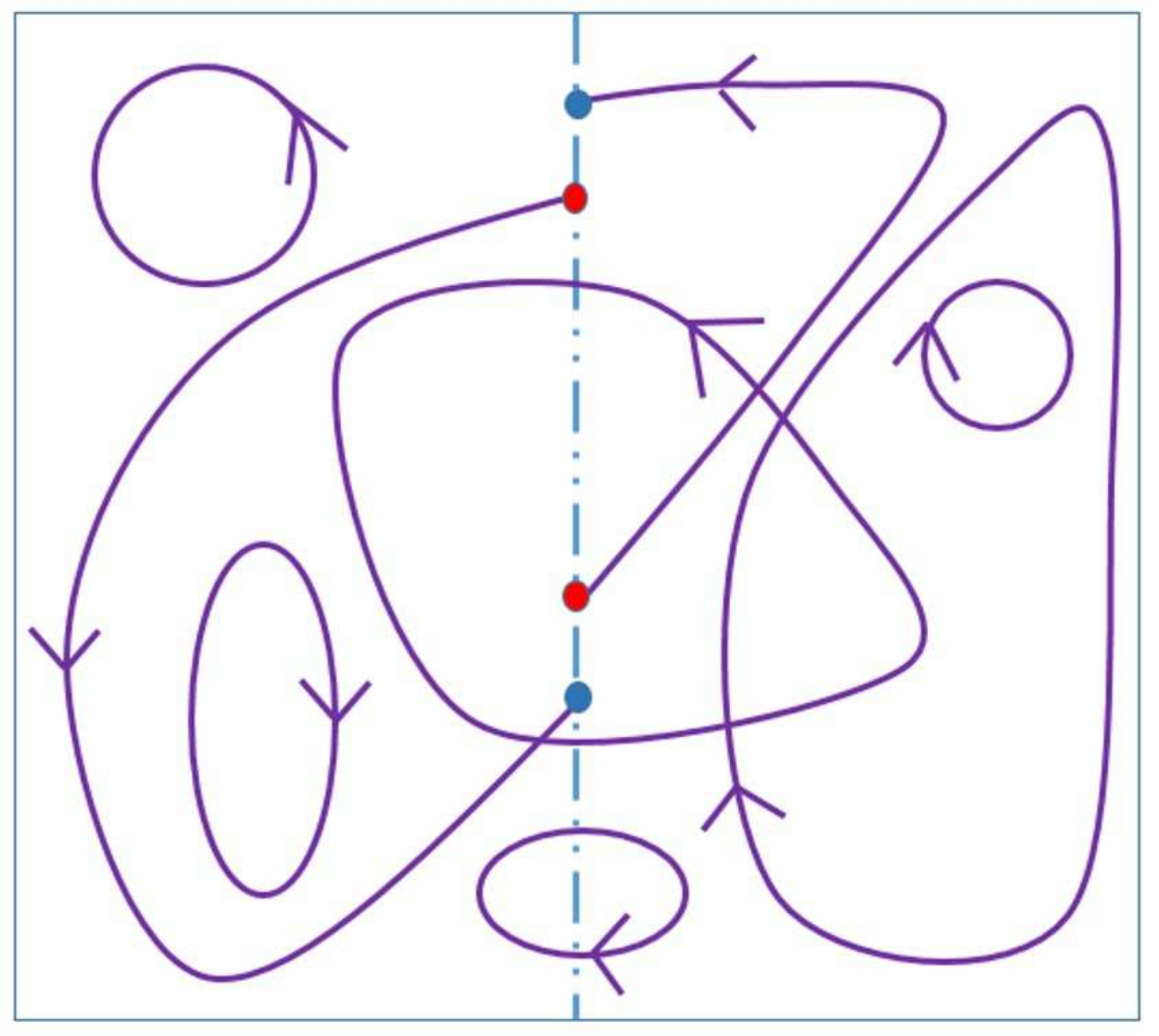}
 \caption{To have a tensor product Hilbert space, we must also include states with open string configurations, $i.e.$ charges.}
 \label{fig:open}
 \end{minipage}
 \end{figure}

Before describing these arguments in a bit more detail, the single most important issue to address is how precisely to define entanglement in a gauge theory, which requires some intuition regarding the Hilbert space of the theory.  The theory we start from is simply the Maxwell field theory of a $U(1)$ gauge field, which has the following action:
\begin{equation}
S = \int d^4x \,F^{\mu\nu}F_{\mu\nu}
\label{maxwellact}
\end{equation}
where the field strength is $F_{\mu\nu} = \partial_\mu A_\nu - \partial_\nu A_\mu$.  Despite the simplicity of the model, defining a notion of entanglement in this theory is a bit tricky.  In order to define entanglement, one must first have a notion of spatially partitioning degrees of freedom.  More technically, one must have a tensor product decomposition of the Hilbert space.  This presents a bit of a problem for a gauge theory.  If one demands that we consider our Hilbert space to only consist of physical photon states, then there is no such tensor product description.  The restrictions of gauge invariance are too stringent for us to independently vary the states of the two halves of any partition.

This problem, and its resolution, are most easily understood geometrically.  By formulating the theory directly in terms of the electric field (see \cite{me}, for a review), Maxwell theory can be viewed as a quantum mechanical theory of closed electric field lines.  These one-dimensional ``strings" serve as the fundamental dynamical variables of the gauge theory, forming a condensate in the deconfined phase\cite{xiao}.  The ground state wavefunction can be written as a superposition of closed field line configurations, such as seen in Figure \ref{fig:closed}.  This restriction to closed loops is precisely the obstruction to decomposing the Hilbert space into a tensor product structure.  The closed loop constraint enforces strict boundary conditions on any bipartition of the system: If a loop passes into the boundary on one side, it must come out of the boundary on the other side.  A loop cannot simply end at the boundary.  This restriction on boundary conditions prevents us from independently varying the states on the two sides of the partition.  A tensor product Hilbert space, on the other hand, would necessarily contain states with mismatched boundary conditions.  For a system of strings, mismatched boundary conditions correspond to states with field lines which end on the boundary, as seen in Figure \ref{fig:open}.

In the early days of studying the gauge theory entanglement problem, the primary technique for dealing with the tensor product issue was a formal algebraic procedure\cite{casini}, which we will comment on further in Section 4.  However, in light of the geometric picture described above, the simplest resolution becomes fairly obvious:  In order to obtain a tensor product Hilbert space, and thereby define entanglement, one may simply work in the ``extended" Hilbert space consisting of both closed and open strings\cite{buiv1,buiv2,donabelian,don1,kuo,ghosh,aoki,don2,radi,me,aspect,zuo,trivedi,lin}.  By embedding the pure photon states into this larger Hilbert space, one obtains a straightforward definition of entanglement.  Furthermore, this notion of gauge theory entanglement matches up precisely with that already in wide use in the condensed matter literature, in the context of topological phases of matter\cite{kitaev,levin}.

But how do we know that this definition is physically meaningful?  Importantly, the extended Hilbert space has a direct physical interpretation.  An endpoint of an electric field line is a point where $\vec{\nabla}\cdot\vec{E}$ is nonzero, which simply corresponds to a charge of the gauge theory.  Indeed, in condensed matter treatments of gauge theories, it is common to \emph{define} charges as the endpoints of strings.  The mass of charges corresponds to the energy cost associated with opening a string, and charge dynamics are inherited directly from the dynamics of open strings.  More formally, in going to the extended Hilbert space, we should generalize our theory to:
\begin{equation}
S = \int d^4x\,( F^{\mu\nu}F_{\mu\nu} + \mathcal{L}_{m} + J^\mu A_\mu)
\label{matteract}
\end{equation}
where $\mathcal{L}_m$ is a Lagrangian describing the endpoints of open strings ($i.e.$ charges) with mass $m$, and $J^\mu$ is the associated current.  In the limit $m\rightarrow\infty$, Maxwell theory should provide an appropriate low-energy description of the problem for most purposes, corresponding to a theory of closed strings.  Importantly, however, the action of Equation \ref{matteract} admits a tensor product decomposition, whereas the action of Equation \ref{maxwellact} did not.  In field theory language, the addition of these charged matter fields in the action may appear $ad$ $hoc$.  But the geometric picture of Maxwell theory as a theory of strings makes it clear that such fields play a natural role.  The crucial lesson is that gauge fields and their corresponding charged particles are not really independent objects, but rather are two sides of the same coin.  Both can be understood within the unified framework of a dynamical theory of strings: photons as the transverse fluctuations of strings, and charged particles as string endpoints.  Furthermore, recent arguments have indicated that such a close relationship between gauge fields and charges is mandated by quantum gravitational effects, which require that gauge fields should be emergent.\cite{harlow}

\section{Calculation of Entanglement Entropy}

With the intuition for the gauge theory's Hilbert space in hand, we now summarize some of the main techniques and results that have appeared in the literature in recent years for the entanglement entropy of Maxwell theory.  We focus here on the methods which we consider most physical.  We will comment on some alternative perspectives in the following section.

\subsection{The Conformal Field Theory Method}

The earliest ideas for calculating the entanglement entropy stemmed from the fact that pure Maxwell theory in $(3+1)$ dimensions is a conformal field theory.  Such a CFT is characterized by two central charges, $a$ and $c$ (following the treatment of Reference \cite{rt}).  These central charges are defined by the trace anomaly of the CFT, relating the expectation value of the trace of the stress-energy tensor to the curvature of spacetime\cite{duff}:
\begin{equation}
\langle T^\mu_{\,\,\,\mu}\rangle = -\frac{c}{8\pi}W_{\mu\nu\rho\sigma}W^{\mu\nu\rho\sigma} + \frac{a}{8\pi}\tilde{R}_{\mu\nu\rho\sigma}\tilde{R}^{\mu\nu\rho\sigma}
\label{curve}
\end{equation}
where $W$ and $\tilde{R}$ are the Weyl tensor and the dual curvature tensor, respectively.  (We refer the reader to Reference \cite{rt} for details on the precise forms of these curvature tensors, which will not be important here.)  Since the trace anomaly is only nontrivial in a curved spacetime, one may wonder why it has anything to do with calculating entanglement entropy in flat spacetime, which is our primary interest.  The answer comes from the replica method, a very powerful tool in the study of entanglement.  While the entanglement entropy, $S = -Tr[\rho\log\rho]$, is the desired end goal of our calculation, it is often useful to first calculate $Tr[\rho^n]$ for integers $n>1$, which is closely related to the Renyi entropies.  The entanglement entropy can then be obtained by analytically continuing to non-integer $n$ and taking the following derivative:
\begin{equation}
S = -\frac{\partial}{\partial n}\log Tr[\rho^n]|_{n=1}
\end{equation}
The advantage here is that the Renyi entropies are typically much easier to calculate than the entanglement entropy, due to a convenient geometric representation of ground state wavefunctions, the logic of which we briefly recapitulate.

Ground state wavefunctions are easy to isolate.  Starting with a generic wavefunction, which in general has nonzero overlap with the ground state, we can pick out the ground state by simply evolving in imaginary time.  After evolving for imaginary time $\tau$, any state with energy $E$ will be weighted down by a factor of $e^{-E\tau}$.  At large $\tau$, only the true ground state will remain.  Using this logic, we can write the ground state wavefunction in terms of an imaginary time path integral which evolves a state over half of Euclidean spacetime:
\begin{equation}
\Psi[A_\mu] \propto \int \mathcal{D}A'_\mu e^{\int_{\tau>0} d^3xd\tau \mathcal{L(A'_\mu)}}
\end{equation}
with the important boundary condition that $A'_\mu(\tau=0) = A_\mu$.  The conjugate of this wavefunction, $\Psi^\dagger[A_\mu]$, can be written as a path integral over the other half of the spacetime:
\begin{equation}
\Psi^\dagger[A_\mu] \propto \int \mathcal{D}A'_\mu e^{\int_{\tau<0} d^3xd\tau \mathcal{L(A'_\mu)}}
\end{equation}
The reduced density matrix for a spatial region A can then be constructed by gluing these path integrals together ($i.e.$ matching boundary conditions) along the complementary region B.  The resulting density matrix is given by an imaginary time path integral over all of Euclidean spacetime:
\begin{equation}
\rho_A[A_\mu,A'_\mu] \propto \int \mathcal{D}A''_\mu e^{\int d^3xd\tau \mathcal{L}(A''_\mu)}
\end{equation}
subject to a branch cut boundary condition along region A: $A''_\mu(\tau = 0^+,x\in $A$) = A_\mu$ and $A''_\mu(\tau=0^-,x\in $A$)=A'_\mu$.  Powers of $\rho$ can be formed in a similar fashion.  To form $\rho^2$, we simply take two copies of this path integral and glue them together, matching the top half of one branch cut to the bottom half of the other, and vice versa.  The result is a path integral over a 2-sheeted manifold.  Similarly, $\rho^n$ is obtained as a path integral over an $n$-sheeted manifold, $M_n$.  Taking the trace, $Tr[\rho^n]$, we obtain the corresponding partition function, $Z_n$, on the $n$-sheeted manifold, up to a normalization constant, $(Tr[\rho])^{-1} = Z_1^{-1}$.  The entanglement entropy is then given by:
\begin{equation}
S = -\frac{\partial}{\partial n}\log \bigg(\frac{Z_n}{Z_1}\bigg)\bigg|_{n=1}
\end{equation}
Up to now, the analysis has been completely general, applying to any local quantum field theory.  Now, however, we specialize to the case of a conformal field theory, which has no intrinsic length scale.  As such, changing the partition size $L$ must be equivalent to a corresponding rescaling of the metric $g_{\mu\nu}$ of spacetime.  To extract how the entanglement entropy scales with partition size, we can simply take an appropriate derivative with respect to the metric:
\begin{equation}
L\frac{\partial}{\partial L}S = -\frac{\partial}{\partial n}\int d^4x g^{\mu\nu}\frac{\delta}{\delta g^{\mu\nu}} (\log(Z_n)-\log(Z_1))\bigg|_{n=1} = -\frac{\partial}{\partial n}\int_{M_n} d^4x \langle T^\mu_{\,\,\,\mu}\rangle\bigg|_{n=1}
\end{equation}
The final integral is over the $n$-sheeted manifold $M_n$.  We have used the definition of the stress-energy tensor as the derivative of the action with respect to metric, and we have taken advantage of the fact that $\langle T^\mu_{\,\,\,\mu}\rangle$ vanishes in flat spacetime, so there is no contribution from the $M_1$ term.  This expectation value can now be related to the curvature of $M_n$ via the conformal anomaly (Equation \ref{curve}).  Note that $M_n$ is flat away from the conical singularity at the partitioning surface, so we will be left with only an integral over this surface.  Indeed, after some herculean geometric manipulations\cite{rt}, one can extract the desired subleading logarithm in the entanglement entropy, with coefficient $\gamma$ given by an integral over the partitioning surface ($\Sigma$):
\begin{equation}
\gamma = \frac{a}{180}\bigg(\int_\Sigma d^2x \sqrt{h} \frac{K}{2\pi}\bigg) + \frac{c}{240\pi}\bigg(\int_\Sigma d^2x \sqrt{h} (K_{ab}K^{ab} - \frac{1}{2}K^2)\bigg)
\label{surface}
\end{equation}
where $h$ is the induced metric on the partitioning surface and $K_{ab}$ is its curvature.  The quantities $a$ and $c$ are the central charges defined in Equation \ref{curve}, which for a Maxwell field theory take the values $a=31\pi/90$, $c=\pi/5$ (adjusting the results of \cite{freedman} to fit our normalization scheme).  For simplicity, most authors focus on the case of a spherical partitioning surface, for which the second integral vanishes, and the first integral yields the simple answer:
\begin{equation}
\gamma_{sphere} = \frac{31}{45}
\end{equation}
However, focusing on the sphere has the unfortunate side effect of de-emphasizing a crucial aspect of Equation \ref{surface}: It is given by an integral of a local quantity over the partitioning surface.  We will return to this important fact shortly.

\subsection{The Thermodynamic Method: Pure Gauge Theory}

Slightly after the CFT arguments were put forward, the entanglement entropy for Maxwell theory with a spherical partition was calculated by Dowker, using a different method, and seemingly yielding a different result\cite{dowker}.  Making use of the geometric representation of the wavefunction from the previous section, along with some clever conformal mapping, one obtains a simple explicit expression for the reduced density matrix of any CFT.  Let the Hamiltonian density of our theory be $\mathcal{H}$, such that the Hamiltonian is $H = \int d^3x \,\mathcal{H}$.  Then the reduced density matrix for the interior of a sphere of radius $R$ is given by \cite{chm}:
\begin{equation}
\rho\propto \exp\bigg\{-\pi\int_{r<R}d^3x\,\bigg(\frac{R^2-r^2}{R}\bigg)\mathcal{H}\bigg\}
\end{equation}
(The above equation is written for three spatial dimensions, but similar results hold in any dimension.)  This expression looks very much like a thermal density matrix, $e^{- \int d^3x\,\mathcal{H}/T}$, but with a position-dependent temperature, diverging as $T(r) \sim 1/(R-r)$ near the boundary.  We will return to this point of view in the next section.  Alternatively, this density matrix can be interpreted as a uniform-temperature thermal density matrix in de Sitter spacetime, with the prefactor of $\mathcal{H}$ representing curvature.  The entanglement entropy is then given by the thermal entropy of our original theory defined on this curved spacetime.  By a direct calculation of photon thermodynamics in de Sitter spacetime, Dowker calculated a value of $\gamma = 16/45$ for the logarithmic coefficient of Maxwell theory with spherical partitioning surface\cite{dowker}.

The disagreement of this result with the trace anomaly was troubling at first, but work by several independent groups has managed to shed some light on the discrepancy\cite{don1,don2,kuo,zuo,trivedi}.  Dowker's work focused on the thermodynamic contribution from the bulk photon degrees of freedom.  However, one must also account for the need to match boundary conditions between the two sides of the partition (see Figure \ref{fig:closed}).  There is an extra entropy associated with different choices of boundary conditions on the partitioning surface, corresponding to different electric flux configurations through the boundary.  In the language of the conformal mapping, these can be interpreted as charged sources living on the boundary of de Sitter space.  Thus, there should actually be two independent contributions in the thermodynamic calculation of entropy, one coming from bulk photons and one coming from boundary charges.  The charge contribution takes the form of a Shannon entropy for the positions of charges on the boundary.  If the flux on the boundary has a probability distribution $p(E)$, then the corresponding entropy is:
\begin{equation}
S_{boundary} = -\int \mathcal{D}E\, p(E)\log p(E)
\end{equation}
where the integral runs over all possible configurations of the electric flux on the boundary (which is restricted to zero net flux, since there are no charges in the bulk).  For pure Maxwell theory, the action is quadratic, so the probability distribution for $E$ may be determined exactly:
\begin{equation}
p(E) \propto e^{-\frac{1}{2}\int dxdyE(x)E(y)G^{-1}(x,y)}
\end{equation}
where $G(x,y) = \langle E(x)E(y)\rangle$.  By direct calculation\cite{trivedi}, it can be found that the corresponding entropy for a spherical partition gives a contribution of $1/3$ to the logarithmic coefficient $\gamma$.  After accounting for both the bulk and boundary contributions, we therefore obtain that the logarithmic coefficient for a spherical partition is:
\begin{equation}
\gamma = \frac{16}{45} + \frac{1}{3} = \frac{31}{45}
\end{equation}
in agreement with the trace anomaly.  For strictly pure Maxwell theory, the trace anomaly seems to have the final say on the logarithmic coefficient.

\subsection{The Thermodynamic Method: Dynamical Charges}

However, this is not quite the end of the story.  While these issues were being sorted out in the high energy literature, a similar calculation was performed in a condensed matter context by the present author and T. Senthil, obtaining a slightly different result.  In condensed matter systems, Maxwell theory arises as a low-energy limit of a theory which also possesses gapped charges.  We therefore kept dynamical charges in the problem, with a large (but finite) mass.  In the presence of such a mass scale, the field theory is no longer conformal, so the trace anomaly method and thermal de Sitter method cannot be rigorously used.  But despite being no longer conformal, we can still take our theory to be relativistic.\footnote{In the context of a $U(1)$ spin liquid, the Lorentz symmetry will be in terms of an emergent ``speed of light" determined by the microscopic Hamiltonian.}  Luckily, there is a similar thermodynamic method which relies only on Lorentz invariance.  In axiomatic QFT circles, this result is known as the Bisognano-Wichmann (BW) theorem\cite{wichmann,bianchi,faulk}.  Suppose we partition our system into two half-spaces, $x_1<0$ and $x_1>0$.  After tracing out the $x_1<0$ region, the reduced density matrix for the remaining half-space will be given by:
\begin{equation}
\rho \propto \exp\bigg\{-\int_{x_1>0}d^dx\,(2\pi x_1\mathcal{H})\bigg\}
\label{bw}
\end{equation}
where $\mathcal{H}$ is the original Hamiltonian density.  (We work in units such that the speed of light is $1$.)  This density matrix has the form of a thermodynamic ensemble, but with a locally defined temperature\cite{wong}:
\begin{equation}
T(x) = \frac{1}{2\pi x_1}
\end{equation}
This invokes a picture where the system is extremely hot at the boundary but cools off to zero temperature in the bulk.  In fact, in certain cases, the exact entanglement entropy can be found simply by integrating the local thermal entropy, $S_{ent} = \int_{x_1>0} S_{th}(T(x))$ \cite{wong,me}.

This local thermal viewpoint allows us to identify two independent contributions to the entanglement entropy.  The photons of the system are gapless and are therefore excited at arbitrarily low temperatures, indicating that the photon distribution extends well into the bulk.  The charge degrees of freedom, on the other hand, have a large mass $m$.  Charges are only excited in nontrivial quantities at temperatures at or above the mass scale.  Thus, thermal charges only exist in a narrow boundary layer of size $m^{-1}$ near the partitioning cut.  In the limit of large mass, the particles are essentially confined to motion only along the boundary.  Just as in the previous thermodynamic method, we thereby obtain two distinct contributions to the entanglement entropy: a bulk contribution from photons and a boundary contribution from charges.  (It is not immediately obvious that these two contributions are totally independent, since while the charges are restricted to the boundary, their attached field lines extend into the bulk.  $A$ $priori$, it is not clear that we can write down two decoupled partition functions.  The crucial insight will come from the fact that charges are screened, a fact which we come to shortly.)

\begin{figure}[t!]
 \centering
 \includegraphics[scale=0.35]{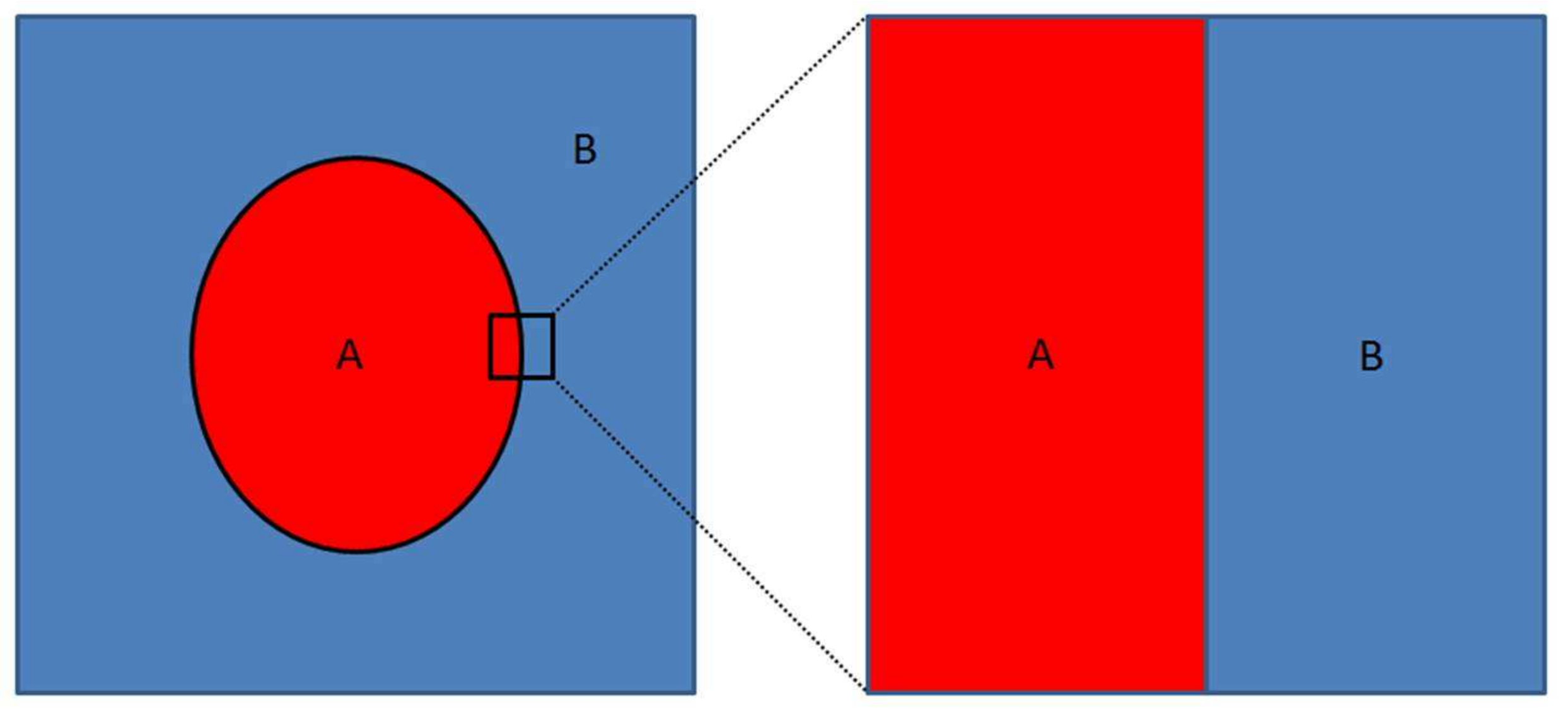}
 \caption{For the charge sector, we can apply the Bisognano-Wichmann result to a generic partition by zooming in on a small patch (which looks locally planar), then summing the contributions from all such patches.}
 \label{fig:planar}
 \end{figure}

We analyze each contribution in turn.  First, we consider the thermal distribution of photons extending into the bulk.  For a flat partition into two half-spaces, the temperature profile for these photons falls off as $T(x)\sim 1/x_1$, where $x_1$ is the distance from the boundary.  However, what we actually want to calculate is the entanglement entropy for some closed partitioning surface.  For simplicity, we focus on the case of a sphere.  We now take advantage of the fact that the photon sector of the theory is still scale-invariant, allowing us to take advantage of the conformal field theory techniques encountered earlier.  We can conformally map the half-space with temperature profile $T(x)\sim 1/x_1$ onto a sphere of radius $R$ with radial temperature profile $T(r)\sim 1/(R-r)$, which is exactly the thermodynamic problem considered by Dowker.  As such, the photon contribution to the entanglement entropy for a spherical partition will be given by the Dowker result:
\begin{equation}
\gamma_{photon} = 16/45
\end{equation}
which is exactly the same bulk contribution considered in the high energy literature.

Besides the thermal photons, we must also account for the entropy of thermally excited charges on the boundary.  Since this charge gas possesses a natural scale, given by the charge mass $m$, we can no longer use conformal mapping techniques to transform the flat-space BW theorem into a result applicable to curved surfaces.  However, we can still obtain the entanglement entropy for an arbitrary surface when the charge mass is very large.  As argued in the past\cite{bianchi,me}, as long as the curvature is small relative to the relevant length scales in the problem, we can zoom in on individual patches of a generic partition, which look locally planar, then sum the contribution to entanglement entropy from the different patches, as in Figure \ref{fig:planar}.  In the present context, let $\mathcal{R}$ be the radius of curvature of the surface.  We then need $m\mathcal{R}\gg 1$, such that the surface looks flat on distance scales of order $1/m$.  As $m$ becomes very large, this approximation will hold for partitioning surfaces of almost any curvature.  This intuition has been put on more rigorous footing in the context of the AdS/CFT correspondence \cite{faulk}, in which it was shown that the entanglement Hamiltonian for an arbitrary partition approaches that of the half-space for large entanglement energies, further justifying the approximations made here.  Assuming that we consider surfaces only with $\mathcal{R}\gg 1/m$, we can safely work within a patch framework, in which the entanglement entropy is completely independent of curvature.

We now must calculate the entanglement entropy of the thermal boundary Coulomb gas, which importantly is subject to a neutrality constraint.  This constraint arises from the fact that there are no charges in the bulk, so there is zero net electric flux through the surface.  Furthermore, since these charges interact through a three-dimensional $1/r$ Coulomb potential, they will always exist in a screened phase.  We must therefore investigate the thermodynamics of a screened Coulomb gas subject to a neutrality constraint.  Precisely such an analysis was carried out in our previous work \cite{me}.  The partition function for such a neutral boundary gas can be written as:
\begin{equation}
Z = \int \mathcal{D}n(x)\,\delta\bigg(\int d^{d-1}x \,n(x)\bigg)P(\{n(x)\})
\end{equation}
where $P(\{n(x)\})$ represents the probability distribution of particle number $n(x)$.  We now take advantage of the fact that the Coulomb gas is screened, such that only short-range correlations are present in the system, out to a distance of the screening length $\lambda$.  We now coarse-grain space into patches of linear size $\lambda$, such that the probability distribution approximately factors into local distributions in each patch, $i.e.$ $P(\{n(x)\}) = \prod_x f(n(x))$, for some probability distribution $f(n(x))$ on the local density.  (We have assumed that all sites are equivalent, such that $f$ takes the same form for all $x$.)  We then take advantage of a Fourier representation of the delta function to write:
\begin{equation}
Z = \int \mathcal{D}n(x)\int db\,e^{ib\int d^{d-1}x\,n(x)}\prod_x f(n(x)) = \int db\prod_x \int dn(x) e^{ibn(x)}f(n(x))
\end{equation}
Integrating over the density, we obtain:
\begin{equation}
Z = \int db \prod_x \tilde{f}(b) = \int db\,(\tilde{f}(b))^{(L/\lambda)^{d-1}}
\end{equation}
where $\tilde{f}(b)$ is the Fourier transform of the local probability distribution.  The product over $x$ raises $\tilde{f}$ to the power $(L/\lambda)^{d-1}$, where $\lambda$ acts as the effective short-distance cutoff.  We now take advantage of the fact that $L/\lambda$ is extremely large, such that the partition function is dominated by the behavior of $\tilde{f}$ near its maxima.  We analyze the case where $\tilde{f}$ has a single maximum, for simplicity (though multiple maxima can be easily handled).  For a generic maximum at $b_0$, $\tilde{f}$ can be expanded to lowest order as $\tilde{f}(b) = c(1-\alpha(b-b_0)^2)$ for constants $c$ and $\alpha$.  In order to simplify the integral, we can replace $\tilde{f}(b)$ with any other function with an equivalent second order Taylor expansion around $b_0$.  A particularly convenient choice is the Gaussian $ce^{-\alpha(b-b_0)^2}$.  Making this replacement, the partition function can be evaluated as:
\begin{equation}
Z \approx \int db\,c^{(L/\lambda)^{d-1}}e^{-\alpha(L/\lambda)^{d-1} (b-b_0)^2} = \frac{c^{(L/\lambda)^{d-1}}}{2\sqrt{\pi\alpha (L/\lambda)^{d-1}}}
\end{equation}
We recall that, in terms of the partition function, the thermal entropy is given by:
\begin{equation}
S = \beta(E-F) = -\beta\partial_\beta \log Z + \log Z = \alpha\bigg(\frac{L}{\lambda}\bigg)^{d-1} - \bigg(\frac{d-1}{2}\bigg)\log(L/\lambda)
\end{equation}
where $\alpha$ is some nonuniversal constant.  The final step is to determine the screening length of the boundary Coulomb gas.  For a two-dimensional gas of particles interacting through a $1/r$ potential ($i.e.$ a \emph{three}-dimensional Coulomb interaction), it is a straightforward exercise to show that the screening length behaves as $\lambda\sim T/n_2$, where $n_2$ is the two-dimensional density ($i.e.$ particles per area) on the boundary.  Since the three-dimensional density behaves as $n_3\sim \epsilon^{-3}$, where $\epsilon$ is the lattice scale, and since the two-dimensional boundary layer has thickness $m^{-1}$, the boundary density scales as $n_2\sim m^{-1}\epsilon^{-3}$.  Since the temperature of the boundary goes all the way up to $T\sim \epsilon^{-1}$, we see that the screening length behaves as $\lambda\sim m\epsilon^2$.  In three spatial dimensions, the charge contribution to the entanglement entropy then becomes:
\begin{equation}
S_{charge} = \alpha(L/m\epsilon^2)^2 - \log(L/m\epsilon^2)
\end{equation}
Note that, while the area law coefficient $\alpha$ is nonuniversal, the subleading logarithm coefficient takes the universal value of $1$.  This makes some intuitive sense, in that the screened boundary gas has only short-range correlations, like a gapped system.  As in a gapped system, the subleading term depends only on the topology of the partitioning surface, not its precise shape.  Specifically, it is straightforward to generalize to a partition with multiple connected components, in which case the charge contribution to entanglement entropy behaves as:
\begin{equation}
S_{charge} = \alpha (L/m\epsilon^2)^2 - b_0 \log (L/m\epsilon^2)
\end{equation}
where $b_0$ is the zeroth Betti number, counting the number of connected components of the partition.  For a connected surface, we can simply write the contribution to the logarithmic coefficient as:
\begin{equation}
\gamma_{charge} = 1
\end{equation}
However, there is an important distinction between this contribution to $\gamma$ and that from the photon sector, in that the present logarithm features the charge mass $m$, whereas the photon logarithm was written entirely in terms of the short-distance cutoff $\epsilon$.  We see that, in the presence of a dimensionful parameter like $m$, our earlier CFT analysis of entanglement entropy breaks down.  Whereas the CFT result relied on $L$ and $\epsilon$ being the only scales in the problem, a finite charge mass $m$ allows for extra terms in the entanglement entropy which can modify the behavior of the logarithm.  Adding this contribution to the result from the bulk photons, our total result for the logarithmic coefficient for a spherical partition is:
\begin{equation}
\gamma = \frac{16}{45} + 1
\end{equation}
This answer has a different boundary contribution from the trace anomaly result, $1$ versus $1/3$.  However, the physical origin of this difference is clear.  The boundary $1/3$ of the trace anomaly result arises from the long-range correlations of an \emph{unscreened} Coulomb gas.  In the presence of dynamical charges, on the other hand, screening leaves only short-range correlations on the boundary, leading to a boundary contribution of $1$.

To summarize the results of this section, the bulk contribution to the entanglement entropy is unambiguously given by a thermal distribution of photons, which for a sphere leads to a logarithmic contribution of $16/45$.  The boundary contribution, however, depends sensitively on the treatment of charges in the theory.  In strictly pure Maxwell theory, the boundary distribution consists of nondynamical test charges, with long-range correlations, resulting in a contribution to $\gamma$ of $1/3$, recovering the trace anomaly result.  However, if one considers a theory with dynamical charges, of large but finite mass, then the boundary will enter a screened phase, without long-range correlation.  In this case, the boundary contribution to the logarithmic coefficient is $1$.

It is important to consider the precise nature of the limits in which the two results hold, as well as the crossover between them.  The condensed matter result relies on having a finite screening length, smaller than the size of the partitioning surface being considered.  In other words, we require $\lambda\sim m\epsilon^2 \ll L$.  This condition is naturally satisfied in the limit $m\epsilon \ll 1$, in which the large mass is held finite while the cutoff scale is taken to zero.  Indeed, since we naturally consider partitions with $L \gg \epsilon$, the condensed matter result will continue to hold even as $m\epsilon$ tends to $1$.  In contrast, the high energy result holds in the case when the theory can be regarded as a pure gauge theory all the way up to the cutoff scale, which corresponds to $m\epsilon\gg 1$, such that dynamical charges do not appear in the spectrum of the theory.  This corresponds to the limit where the screening length is larger than the system size, such that the fields associated with charges appear to be long-ranged.  For a fixed partition size, we expect a fairly rapid transition from the condensed matter to the high energy result around $\lambda \sim L$.

\subsection{Topological Isolation Procedure}

We have now discussed two different answers for the logarithmic coefficient of Maxwell theory: the trace anomaly result from the high energy literature and the topological result from the condensed matter literature.  However, based on the discussion so far, these results have differed primarily at the level of unpleasant fractions.  To prevent the analysis from veering into mystic numerology, we here discuss a much more physical way to distinguish between the two results.

As we discussed in the introduction, not all of the entanglement entropy is actually an interesting quantity.  In fact, the dominant area law term contains almost no useful information about the system whatsoever.  This is because most of the entanglement in a local quantum field theory is short-ranged ($i.e.$ between nearby degrees of freedom).  The area law term arises directly from such short-range entanglement, corresponding to short-range Bell pairs which are cut by the partition.  In contrast, the subleading topological term is a function of the number of connected components of the partitioning surface.  Unlike surface area, the number of connected components has no expression as an integral of a local quantity over the surface.  Rather, it is a global property of the entire object.  This distinction is the key idea behind the concept of topological entanglement entropy \cite{kitaev,levin}, widely used in the study of topological phases of matter, which provides a mechanism to isolate ``global" terms by eliminating contributions from local surface integrals.  The leftover portion of the entanglement entropy should be a topological invariant, capturing only the universal physics.  This same procedure can also be carried over directly to the analysis of Maxwell theory.

\begin{figure}[t!]
 \begin{minipage}[b]{0.45\linewidth}
 \centering
 \includegraphics[scale=0.5]{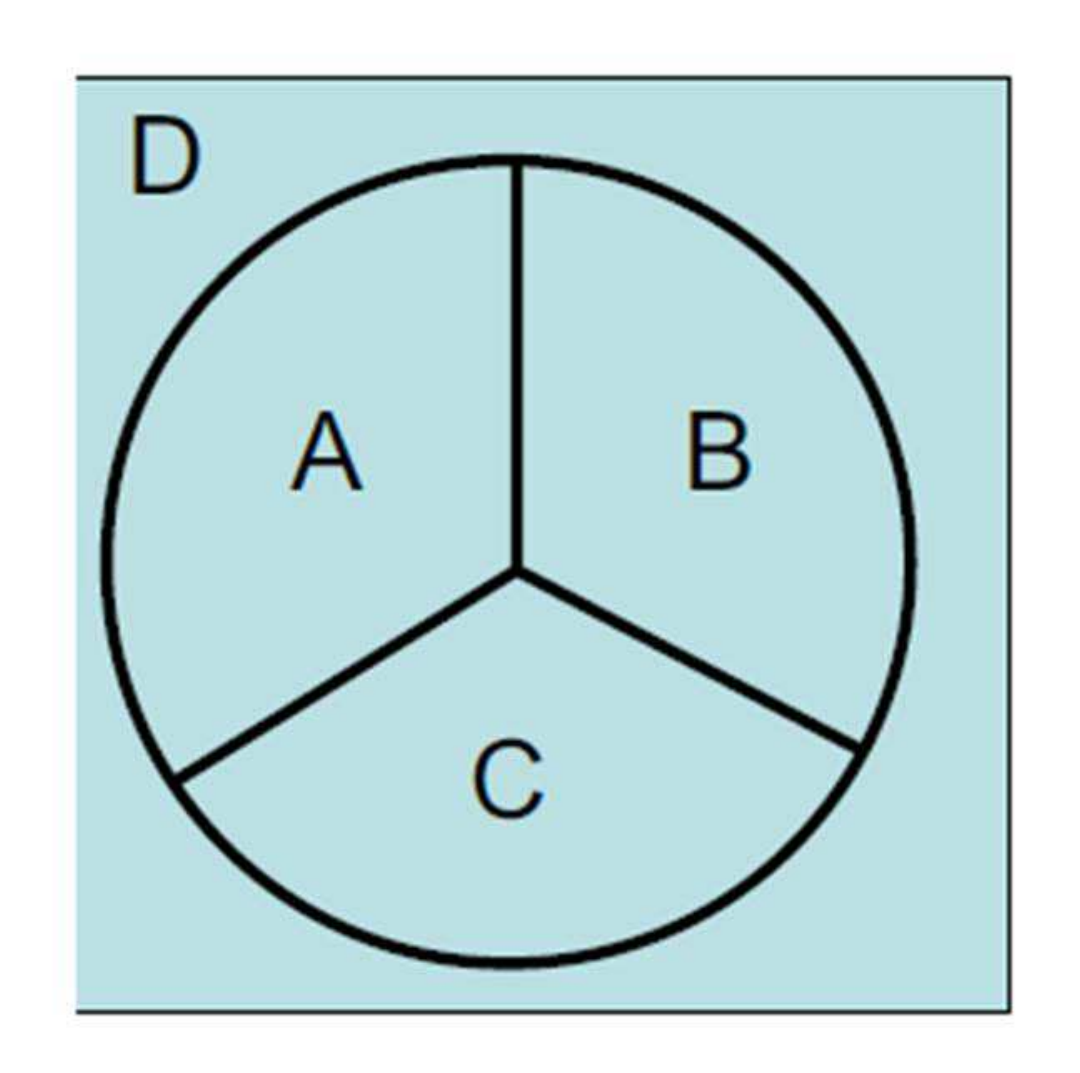}
 \caption{The Kitaev-Preskill construction allows for the isolation of topological entanglement entropy in two dimensions \cite{kitaev}.}
 \label{fig:kitaev}
 \end{minipage}
 \hspace{1cm}
 \begin{minipage}[b]{0.45\linewidth}
 \centering
 \includegraphics[scale=0.5]{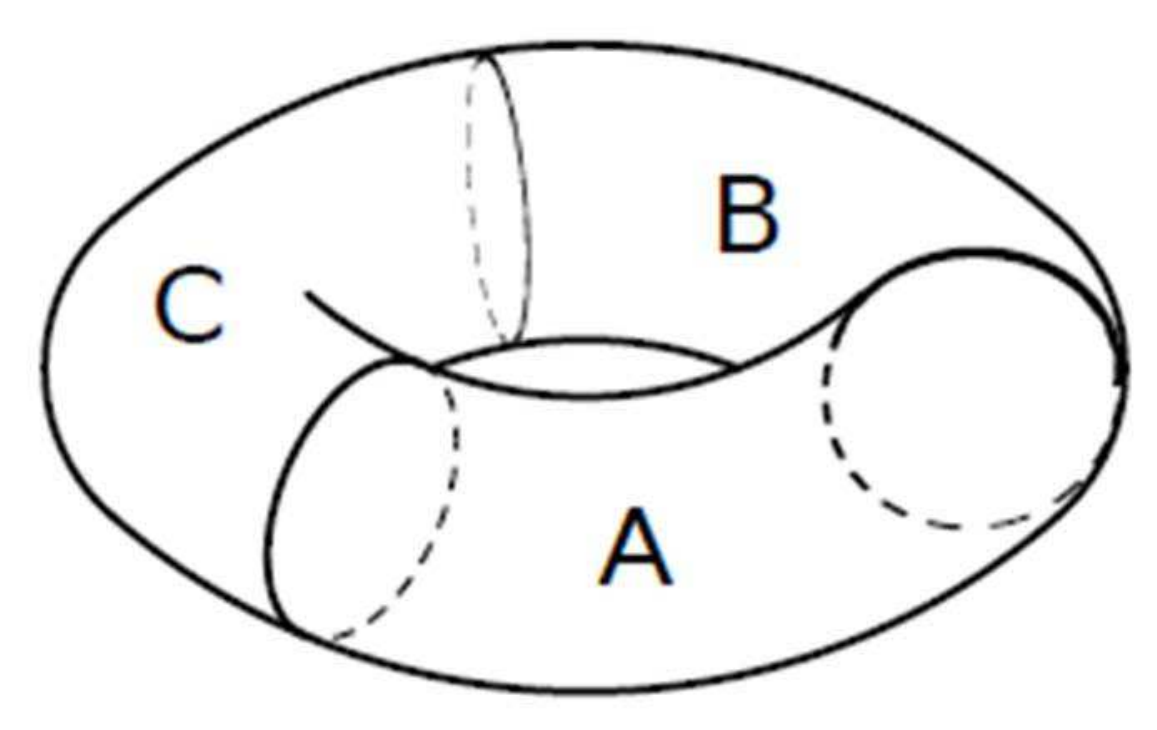}
 \caption{The Grover-Turner-Vishwanath construction generalizes the Kitaev-Preskill scheme to three dimensions and isolates the analogue of topological entanglement entropy in Maxwell theory \cite{grover}.}
 \label{fig:grover}
 \end{minipage}
 \end{figure}

The method proceeds via a geometric construction.  We must carefully choose a set of partitions with partially overlapping surfaces, allowing us to eliminate all surface terms without also killing the topological piece.  In $(2+1)$ dimensions, the simplest such arrangement is the Kitaev-Preskill scheme\cite{kitaev}, depicted in Figure \ref{fig:kitaev}, featuring four distinct regions: A, B, C, and D.  An appropriate $(3+1)$-dimensional analogue, constructed by Grover, Turner, and Vishwanath\cite{grover}, is depicted in Figure \ref{fig:grover}.  In either case, we define $S_A$ to be the entanglement entropy after tracing out everything but region A, and similarly for other regions.  We can then define the topological entanglement entropy as:
\begin{equation}
S_{top} = S_A + S_B + S_C - S_{AB} - S_{AC} - S_{BC} + S_{ABC}
\label{top}
\end{equation}
One can easily check that surface integrals coming from each piece of the boundary will cancel between the various terms, killing any local contributions.  In contrast, contributions from certain topological invariants, such as the zeroth Betti number, will survive this procedure, yielding a nonzero topological entanglement entropy.  For example, in a $\mathbb{Z}_2$ lattice gauge theory, we obtain $S_{top} = -\log 2$, which is precisely the universal piece captured by the effective Chern-Simons description.

Let us now apply this same procedure to the case of Maxwell theory and examine what happens to the logarithmic coefficient.  The trace anomaly calculation yielded a contribution to $\gamma$ which was given entirely by a local surface integral (see Equation \ref{surface}).  Therefore, the trace anomaly result for the logarithm is topologically trivial and will be totally wiped out by the topological isolation procedure.  (In Reference \cite{trivedi}, it was shown that, within the high energy framework, $S_{top}$ is given by a constant, which in fact fails to be topologically invariant, as a consequence of the photon contribution.)  In contrast, let us consider the condensed matter treatment, featuring dynamical charges.  The bulk photon term still arises from local physics and gets eliminated.  The boundary charge term, however, is now proportional to the zeroth Betti number.  It is then easy to see that we have:
\begin{equation}
S_{top} = -\log L
\end{equation}
In other words, we may write $\gamma_{top} = 1$.  We therefore see that the condensed matter calculation, accounting for dynamical charges, yields a topologically protected contribution to the entanglement entropy, while the trace anomaly calculation does not.  In principle, local contributions to the entanglement entropy are observable as well.  However, there are various lattice issues ($e.g.$ corner contributions \cite{corner}) which can wash out such non-topological terms.  Topological entanglement entropy provides a fully robust metric of entanglement which can be extracted by a well-defined numerical procedure.  (Though of course, extracting such a logarithmic term in a three-dimensional system will be no easy numerical feat.)  In this sense, the topological contribution from dynamical charges provides the most robust entanglement-based characterization of Maxwell theory.

\section{Miscellaneous Issues}

We here take a moment to provide a condensed matter perspective on some other common issues in the study of entanglement in gauge theories, including some alternative methods which have been used to define and calculate the entanglement entropy.

\subsection{The Algebraic Formalism}

Some of the early ideas on gauge field entanglement in the high energy literature focused on an algebraic formalism, emphasizing the role of ``algebras with centers" in a gauge theory\cite{casini}.  The argument was that there is a fundamental ambiguity in the definition of entanglement for a gauge field, based on the fact that there is not a unique way of localizing physical observables on the two sides of a partition.  However, as we discuss, placing certain physically reasonable demands on the gauge field (specifically the ability to couple to charges) resolves the ambiguity and picks out a preferred definition of entanglement.

\begin{figure}[b!]
 \centering
 \includegraphics[scale=0.5]{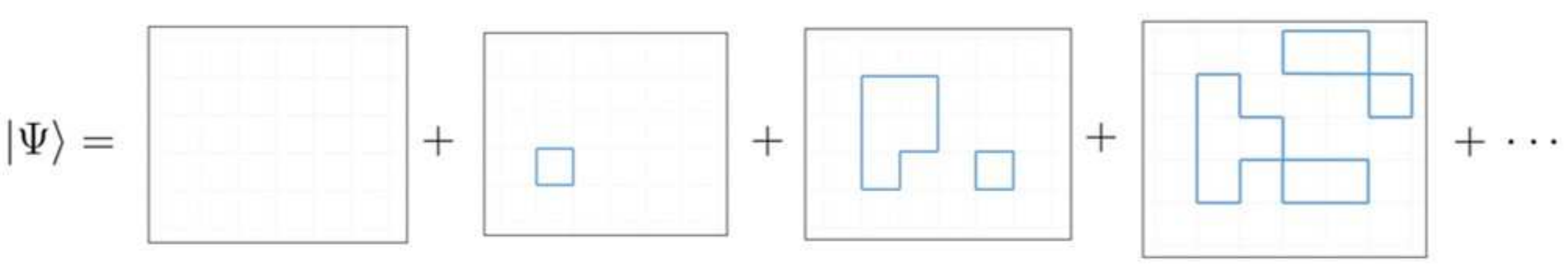}
 \caption{The ground state wavefunction of a $\mathbb{Z}_2$ gauge theory is an equal-weight superposition of all closed string configurations on the links of a lattice.  Equivalently, we can label the state via plaquette variables, in terms of which the wavefunction looks like a direct product state.}
 \label{fig:toric}
 \end{figure}

The difficulty of uniquely localizing observables in a gauge theory relates back to a geometric ambiguity in the way we partition the system.  The essential issue is best illustrated by example.  Consider a simple $\mathbb{Z}_2$ lattice gauge theory, with ground state given by a superposition of all closed loop configurations running along the links of a lattice, as in Figure \ref{fig:toric}.  In all of our discussions in this paper, we have implicitly taken ``partition" to mean a partition of the gauge field itself, defined on the \emph{links} of a lattice.  Equivalently, this can be regarded as partitioning the gauge-invariant ``electric field" of the theory.  Within this definition, the ground state has a nontrivial topological entanglement entropy, specifically $S_{top} = -\log 2$.  But what if we construct an alternative notion of partitioning the system?  For example, instead of labeling a state in terms of links, a system of closed loops in two dimensions can also be described in terms of gauge-invariant plaquette operators.  Specifically, the ground state can be written as:
\begin{equation}
|\Psi\rangle = \prod_{plaquettes} \frac{1}{\sqrt{2}}(|\textrm{loop}\rangle +|\textrm{no loop}\rangle)
\end{equation}
where the product runs over all plaquettes of the lattice.  We see that, if we partition the system in terms of plaquettes, the ground state looks like a direct product state, with zero entanglement.  The notion of entanglement seems to depend sensitively on the way in which we partition the system.  It is precisely this sort of ambiguity which manifests itself in the ``algebra with center" framework, reflecting different ways of localizing the gauge-invariant observables.

So why are we justified in picking the links as the preferred partitioning objects, as we have done throughout this paper, instead of a plaquette representation or something in between?  The answer is that we have demanded that our gauge field be consistently coupled to charged matter fields.  The link representation naturally incorporates charges as the endpoints of open strings.  In contrast, the plaquette representation is only valid within the closed loop sector, breaking down in the presence of charges.  This can be seen, for example, by putting the system on a torus.  The labeling by plaquettes cannot represent a loop wrapped around the torus, so only one of the four degenerate ground states is present in this description.  The plaquette representation is incapable of supporting charge states, whether dynamical or not.

In many physical situations, such as $U(1)$ spin liquids and the Standard Model, dynamical charges are fundamentally a part of the Hilbert space, so we must make the demand that the gauge field should be consistently ``charge-able."  With this stipulation in place, we automatically pick out a preferred choice of partitioning ($i.e.$ a choice of how to localize the gauge-invariant observables).  In particular, the demands of ``charge-ability" automatically require us to work within the framework of the extended Hilbert space.  The ``algebra with center" issues are important to bear in mind for a strictly pure gauge theory.  But as soon as charges are introduced, such concerns immediately evaporate.

\subsection{Entanglement Distillation}

The high energy result for the logarithmic coefficient was given by a sum of two terms:
\begin{equation}
\gamma_{H.E.} = \frac{16}{45}+\frac{1}{3}
\end{equation}
where the first term comes from the bulk photons, and the second arises from behavior at the boundary, which we have interpreted as boundary charges.  Unlike the condensed matter result, which separates into a local piece and a topological contribution, both terms of the high energy result are local.  Nevertheless, there is still a meaningful distinction between them.  The bulk contribution is sometimes referred to as the ``extractable" piece of the entanglement entropy\cite{trivedi}:
\begin{equation}
\gamma_{extract} = \frac{16}{45}
\end{equation}
The logic behind such a separation is that only this part of the entanglement can be removed from the system by operations on the pure photon states, via an ``entanglement distillation" procedure\cite{matter}.  However, it is important to bear in mind that the ``non-extractable" boundary term is far from unphysical.  Indeed, the usual topological entanglement entropy of topological field theories is precisely such a ``non-extractable" boundary term.  As we have discussed, such boundary terms arise from the charge sector of the theory, which is why pure gauge mode operations are insufficient to remove this entanglement from the system.  Indeed, it is precisely the ``non-extractability" of topological entanglement entropy that makes it a robust characterization of a topological phase of matter, since it cannot be removed without driving the system through a phase transition.  Similarly, the boundary contribution to the entanglement entropy of Maxwell theory is more robust than the extractable bulk piece, since modifying the boundary term requires access to the gapped charge degrees of freedom.

\subsection{Gauge-Fixing Procedures}

Another subtle issue which appears widely in the high energy literature is the use of gauge-fixing to simplify calculations.  While gauge-fixing is a perfectly valid procedure if done correctly, loose applications of gauge-fixing can often lead to misleading results.  In particular, it is important to realize that, within the extended Hilbert space formalism, it is no longer valid to apply gauge-fixing only to the gauge field itself, without carefully treating the charge sector of the theory.  For calculating low-energy scattering amplitudes and correlation functions, where charges are largely irrelevant, applying gauge-fixing to the gauge field alone is adequate.  But such a procedure will in general lead to incorrect results for entanglement.  As a trivial example, consider a $\mathbb{Z}_2$ lattice gauge theory in two dimensions.  It is a straightforward textbook exercise\cite{fradkin} to show that applying gauge-fixing to the $\mathbb{Z}_2$ gauge field alone, without accounting for matter fields, can yield a direct product state, with zero entanglement.  This sharply conflicts with the established nontrivial topological entanglement entropy of such a theory.  The seemingly benign act of gauge-fixing can drastically change the entanglement entropy, unless sufficient care is taken to account for charges.

To gain more intuition as to why this type of gauge-fixing fails, let us return to Maxwell theory.  Suppose we start with a generic wavefunction defined in the basis of the electric field, $E$:
\begin{equation}
|\Psi\rangle_{generic} = \int \mathcal{D}E\, f[E]\,|E\rangle
\end{equation}
for some functional $f[E]$.  In order to make sure that this state is in the pure gauge (closed loop) sector of the theory, we insert a delta function into the integral to enforce the Gauss's law constraint, $\nabla\cdot E = 0$, which projects the state into the photon sector:
\begin{equation}
|\Psi\rangle_{photon} = \int \mathcal{D}E\,\delta(\nabla\cdot E)    f[E]\,|E\rangle = \int \mathcal{D}E\mathcal{D}\alpha\, e^{i\int \alpha\nabla\cdot E}f[E]|E\rangle
\end{equation}
where we have introduced an integral representation of the delta function.  We now make a change of basis into the $A$ eigenstates.  The vector potential $A$ is the canonical conjugate variable to $E$, so we have $|E\rangle = \int \mathcal{D}A e^{i\int E\cdot A}|A\rangle$.  Our projected wavefunction then becomes:
\begin{equation}
\begin{split}
|\Psi\rangle_{photon} = \int \mathcal{D}&E\mathcal{D}A\mathcal{D}\alpha\,\,e^{i\int E\cdot (A - \nabla\alpha)}f[E]|A\rangle = \\ \int&\mathcal{D}A\mathcal{D}\alpha \,\,\tilde{f}[A]|A+\nabla\alpha\rangle
\end{split}
\end{equation}
where we have defined the Fourier transform $\tilde{f}[A] = \int\mathcal{D}E \,e^{i\int E\cdot A}f[E]$ and have shifted $A\rightarrow A+\nabla \alpha$ in the last line.  From the above expression, we see that a closed loop wavefunction is forced to be an equal-weight superposition of all gauge-equivalent vector potentials, $A + \nabla\alpha$, for all possible $\alpha$.  It is in this sense that gauge-equivalent configurations are carrying redundant information.  Once we project to the closed-loop sector, $A$ and $A+\nabla\alpha$ will correspond to the same quantum state.  Importantly, however, the extended Hilbert space also includes states where $\nabla\cdot E\neq 0$.  For a state with charges, it is not correct to simply identify $A$ with $A+\nabla\alpha$, as there is also nontrivial transformation within the particle sector.  The configurations $A$ and $A+\nabla\alpha$ are still redundant at the level of the ground state wavefunction, which has no charges, but they are not identical within the full Hilbert space, which includes charged states.

\section{Conclusion}

In this paper, we have reviewed the various calculations of entanglement entropy for the Maxwell theory of a $U(1)$ gauge field, from both the high energy and condensed matter communities.  We have analyzed the discrepancy between the locally-determined trace anomaly result of the high energy literature and the topological contribution from the condensed matter calculation.  We have argued that both results are valid in very slightly different cases.  In the strict limit of pure Maxwell theory, the trace anomaly result is appropriate.  However, for a system with dynamical charges, even highly massive ones, the entanglement entropy undergoes a sudden change, resulting in a topological contribution which is not captured by the conformal field theory calculation.  For most condensed matter purposes, such as the study of $U(1)$ spin liquids, and for many high energy purposes, such as the Standard Model, dynamical charges are fundamentally a part of the Hilbert space, and such topological contributions to the entanglement entropy must be accounted for.

\section*{Acknowledgments}

I would like to acknowledge useful conversations with T. Senthil, Rahul Nandkishore, Leo Radzihovsky, and Mike Hermele.  This work was partially supported by NSF grant PHY 1734006 and partially by Simons Investigator Awards to Leo Radzihovsky and T. Senthil.

\end{document}